\documentclass{sigchi}
\usepackage{balance}  % to better equalize the last page
\usepackage{graphics} % for EPS, load graphicx instead
\usepackage{times}    % comment if you want LaTeX's default font
\usepackage{url}      % llt: nicely formatted URLs

\makeatletter
\def\url@leostyle{%
  \@ifundefined{selectfont}{\def\UrlFont{\sf}}{\def\UrlFont{\small\bf\ttfamily}}}
\makeatother
\def\pprw{8.5in}
\def\pprh{11in}

\usepackage[pdftex]{hyperref}
        {\begin{itemize}[#1,leftmargin=*,parsep=0pt,itemsep=0pt,topsep=0pt,partopsep=0pt]}
        {\end{itemize}}
\hypersetup{
pdftitle={SIGCHI Conference Proceedings Format},
pdfauthor={LaTeX},
pdfkeywords={SIGCHI, proceedings, archival format},
bookmarksnumbered,
pdfstartview={FitH},
colorlinks,
citecolor=black,
filecolor=black,
linkcolor=black,
urlcolor=black,
breaklinks=true,
}

\begin{document}
\newcommand{\tab}[1]{\hspace{..1}\rlap{#1}}
\title{Participatory Militias: An Analysis of an Armed Movement's Online Audience}

\numberofauthors{2}
\author{
  \alignauthor Saiph Savage \\
    \affaddr{Universidad Nacional Aut\'{o}noma de M\'{e}xico \&}\\
   \affaddr{UC Santa Barbara}\\
    \email{saiph@cs.ucsb.edu}\\
    %\affaddr{Optional phone number}
  \alignauthor Andr\'{e}s Monroy-Hern\'{a}ndez\\
    \affaddr{Microsoft Research}\\
    %\affaddr{Address}\\
    \email{amh@microsoft.com}\\
    %\affaddr{Optional phone number}    
%  \alignauthor 3rd Author Name\\
   % \affaddr{Affiliation}\\
    %\affaddr{Address}\\
    %\email{e-mail address}\\
    %\affaddr{Optional phone number}
}

% Teaser figure can go here
%\teaser{
%  \centering
%  \includegraphics{Figure1}
%  \caption{Teaser Image}
%  \label{fig:teaser}
%}

\maketitle

\begin{abstract}
Armed groups of civilians known as ``self-defense forces'' have ousted the powerful Knights Templar drug cartel from several towns in Michoac\'{a}n. This militia uprising has unfolded on social media, particularly in the ``VXM'' (``Valor por Michoac\'{a}n,'' Spanish for ``Courage for Michoac\'{a}n'') Facebook page, gathering more than 170,000 fans. Previous work on the Drug War has documented the use of social media for real-time reports of violent clashes. However, VXM goes one step further by taking on a pro-militia propagandist role, engaging in two-way communication with its audience. This paper presents a descriptive analysis of VXM and its audience. We examined nine months of posts, from VXM's inception until May 2014, totaling 6,000 posts by VXM administrators and more than 108,000 comments from its audience. We describe the main conversation themes, post frequency and relationships with offline events and public figures. We also characterize the behavior of VXM's most active audience members. 
Our work illustrates VXM's online mobilization strategies, and how its audience takes part in defining the narrative of this armed conflict. We conclude by discussing possible applications of our findings for the design of future communication technologies.

\end{abstract}

%\category{H.5.3.}{ Group and Organization Interfaces; Asynchronous 
%interaction}{ Web-based interaction.}

\section{Introduction}
Militias have historically been defined as a group of three or more citizens that gather together in defense of their community, its territory, property or laws~\cite{sumner1823inquiry}. As social media adoption spreads, militias have started to make use of these networked platforms for organizational purposes \cite{crothers2004rage,weeber2003militias,huimas}. From militia groups in Somalia to right-wing militias in the US, social media has become part of militias' communication strategy ~\cite{reclutamiento,weeber2003militias,crothers2004rage,huimas}. %Weeber {et al.}, 
For example, previous work has found that militias in the US use internet technologies because they are user-friendly, reliable, and free~\cite{weeber2003militias}.  

In this paper, we focus on the relationship between militia-related groups on social media and their audiences. Previous work on militia communication practices has identified that some of the goals of these groups when engaging in public channels are to popularize their ideas, distribute propaganda, recruit new militants, and even raise funds~\cite{reclutamiento,weeber2003militias}. With that in mind, we examine which ideas are most ``popular,'' and which posts ``recruit'' participants by engaging them in conversation. We guide our exploration of militia groups on social media with the following research questions:

\begin{enumerate}
\item What content do these groups share online, and how do their audiences react to such content?
\item What content attracts new participants?
\item What are common attributes of the most active people in these online spaces?
\item What are the characteristics of the most popular content?
\end{enumerate}

We explore these research questions using nine months of data from a Facebook page titled 
\emph{``Valor por Michoac\'{a}n SDR\footnote{\url{https://www.facebook.com/ValorPorMichoacan}}''}(VXM), which translates as ``Courage for Michoac\'{a}n.'' This page was created to inform people of ``unsafe situations'' or ``situaciones de riesgo'' (SDR) in the state of Michoac\'{a}n. This type of activity has been reported in previous work that examined how residents of communities afflicted by violence used Twitter to form alert networks and help one another identify potential danger on the streets~\cite{Monroy-Hernandez:2013:NWC:2441776.2441938,de2014narco}. However, shortly after its creation, VXM began to focus more on reporting about the activities of an armed group in the region, the self-defense forces~\cite{VXM}. 

The movement of the self-defense forces began in 2011 when a group of vigilantes from the town of Cher\'{a}n in Michoac\'{a}n expelled a band of illegal loggers backed by the drug cartel of ``La Familia'' \cite{arboles}. This uprising became one of the first cases of ordinary citizens joining forces to challenge the powerful drug cartels. Two years later, new armed groups known as the \emph{autodefensas} (self-defense forces) emerged in the same state but much stronger and  better organized ---and, some argued, with backing from wealthy ranchers \cite{limon}. The self-defense forces gained national and international attention after successfully taking over a number of bigger towns previously controlled by the Knights Templar cartel, a spin-off of the La Familia cartel \cite{limon,arboles}. Although the self-defense forces are a decentralized movement composed of multiple smaller groups, each with their own idiosyncrasies, they share the belief that the government is unable to stop the cartels and bring safety to their towns. Thus, they decided to take action and fight the drug cartels themselves \cite{arboles}. Notice that we refer to these self-defense forces as a militia following Sumner's definition of a militia being a group of concerned citizens who: (1) are non-professional fighters; and (2) have entered a combat situation to defend their land~\cite{sumner1823inquiry} .

It is important to mention that although the VXM Facebook page documents the activities of the Michoac\'{a}n self-defense forces, often with access to seemingly privileged information \cite{processoVXM}, the connection between the page and the militias has not been made official. Furthermore, the administrators of the page remain anonymous. News media, however, often refer to the VXM page as the militias' social media presence \cite{VXM,aristeguiVXM}. The page is also usually one of the first to provide on-site reports about the militias' battles ~\cite{processoVXM}.

We characterize the content that VXM shares and its audience's participation using analytical techniques similar to those for examining discourse between politicians and their audiences \cite{van1997discourse,Robertson:2010:OWP:1858974.1858977}. Our aim is to explore the context under which militia-related content gains attention and to investigate the behavioral patterns of VXM's active audience. In particular, we focus on the topics, locations, public figures, and organizations to which the posts or comments refer. We explored topics, organizations, and public figures because they provide a window into people's interests and issues they may want to make more prominent or de-emphasize \cite{Robertson:2010:OWP:1858974.1858977}. For instance, perhaps the most active audience members emphasized certain social aspects of the situation in Michoac\'{a}n. %Additionally, mentions of geographical locations can help to further profile and contextualize people's discourse \cite{van1997discourse}.

We discuss our findings by exploring the factors that have been used previously to interpret the online activity of militias and other social movements \cite{weeber2003militias}. For instance, we examined the role that information and communication technologies play in mobilizing individuals and in helping people narrate or frame their reality \cite{kelly2006protest}.

Through our study, we reveal how VXM and its audience appear to construct a participatory space where they collectively define what are important events and who are public figures. Together VXM and its audience propose solutions to the problems they recognize, and specify what each person can do to transform the situation in Michoac\'{a}n. We also identify the techniques VXM seems to be using to mobilize its audience for offline collective efforts. Our case study highlights how technology is transforming who can frame the narrative of a region and facilitate the mobilization of individuals for collective efforts.

\section{Related Work: social movements framework}
We position our work within the area of social computing and collective action, particularly focusing on social movements. Social movements are defined as organized collective activities which take place outside political structures, but that attempt to transform existing political, economic, or societal structures~\cite{kelly2006protest}. Militias are considered a type of social movement~\cite{crothers2004rage}. 

The literature~\cite{kelly2006protest,mcadam1996comparative} has identified three universal themes recurring across all social movements: mobilizing structures, opportunity structures, and framing processes. We use these themes to interpret our results and further understand what is occurring with VXM. %, while building on previous work. 

{\bf Mobilizing Structures:} mechanisms through which information communication technologies (ICTs) facilitate mobilizing people to promote an ideology or political cause~\cite{kelly2006protest}. For instance, in the ``Arab Spring'' and ``Occupy'' movements, people used social media to mobilize crowds for street rallies or protests. Similarly, Facebook facilitated the mobilization of political activists in Palestine by letting people connect their private life with their political activities~\cite{Wulf:2013:FAW:2470654.2466262}.
%Wulf et al.~\cite{Wulf:2013:FAW:2470654.2466262} uncovered how Facebook by letting people link private life with political activities facilitated mobilizing 
%political activists in Palestine. 
Lastly, platforms such as Turkopticon allow crowdworkers to mobilize and hold virtual employers accountable for their actions~\cite{Irani:2013:TIW:2470654.2470742}.

{\bf Opportunity Structures:} characteristics of a social system that facilitate or hinder the activity of a social movement~\cite{kelly2006protest}. For instance,
 %in the Israel-Palestine conflict, the Israeli policy of not licensing Palestinian 3G providers was an opportunity structure by forcing Palestinian activists to invest more money in providing internet via landlines, and limiting how their movement could operate~\cite{Wulf:2013:FAW:2470654.2466262}. 
online donation campaigns take advantage of the immediate emotions after a disaster or of friendships to harvest higher levels of participation and donations~\cite{mejova2014giving}. 

{\bf Framing Processes:} ways to interpret reality by labeling one's individual experiences within the experiences of society~\cite{dimond2013hollaback,kelly2006protest}. In social movements, frames represent a set of beliefs and meanings that help to legitimate and motivate the actions of the group. Frames usually emerge by negotiating shared meaning.
Diamond et al.~\cite{dimond2013hollaback} found that people used social media to crowd-source framing processes, and collectively create an interpretation of street harassment.

\section{Data collection and analysis}

Using the public Facebook API, we collected nine months of posts, comments, likes, and shares from the VXM page. The data collection took a snapshot of all activity starting from August 14, 2013 to the date of data collection, May 7,  2014. The details of the dataset are shown in Table \ref{table:st}.
\begin{table}
  \centering
  \begin{tabular}{l c}
    \hline
   
   {Posts}&6,901\\
  {VXM Fans }&158,000\\
  Participants (active audience size) &  127,374  \\
  Comments & 108,967\\
  Post Likes & 1,481,008\\
  Reshares & 364,660\\
    \hline
  \end{tabular}
  \caption{VXM Data.}
  \label{table:st}
\end{table}

\subsection{Content Analysis}  

In order to obtain a descriptive assessment of the VXM page content, we extracted common topics, organizations, and public figures referenced in the posts. First, we familiarized ourselves with the VXM page by frequently reading its posts and exchanging notes and observations among the researchers. It is important to note that the researchers are native Spanish speakers originally from M\'{e}xico. 

{\bf Topics.}  We used a grounded theory approach \cite{groundTheory} to identify the main topics in VXM's posts (Table \ref{table:mism}). One researcher started by independently extracting topics from a set of 200 randomly selected posts. Then, the other researcher analyzed the emerging topics and adjusted the list accordingly. Finally, we looked at a set of 500 randomly selected posts and produced a final list of mutually exclusive topics.

We used oDesk\footnote{oDesk is a platform to contract independent professionals from around the world to perform tasks online.\\ Available at: \url{http://www.odesk.com/}} to hire three Spanish-speaking, college-educated people to categorize the VXM posts independently. First, we asked two coders to categorize each of the 6,901 posts using the topics listed in Table \ref{table:mism}. We asked the coders to pick the ``most relevant''  topic for each post. The two coders agreed on 6,080 posts (Cohen's kappa: 0.54). We then asked the third coder to label the 821 posts upon which the first two coders had disagreed. We then used a ``majority rule'' approach to determine the topic for those posts.

{\bf Public Figures and Organizations.} We identified posts that referred to each of the organizations and public figures involved in the conflict, from public officials to cartel members to militia leaders. We manually created a list of full, common, and nicknames and then identified posts that mentioned any of the names. We collected the names from Wikipedia and from \emph{Proceso}\footnote{\emph{Proceso}, is a well-established magazine covering politics and social issues since 1976.  Available at:\\ \url{http://www.proceso.com.mx/}} magazine. We used the Wikipedia articles, \emph{``Caballeros Templarios Knights Templar cartel''} and \emph{``Grupos de Autodefensa Comunitaria Self-Defense Forces,''} written both in English and Spanish, to manually identify all relevant proper names. We did the same with \emph{Proceso} articles. We used \emph{Proceso's} online search to find articles related to the militias using the keywords \emph{autodefensas} and \emph{caballeros templarios.} We went through each of the articles that appeared in the first page of the search results. We then added or merged the alternate names for each person, both shorter and nicknames. For example, \emph{``Estanislao Beltran''} was also known as \emph{``Papa Smurf,''} and \emph{``Jose Manuel Mireles Valverde''} was usually referred to as just \emph{``Mireles.''}

\begin{figure*}
\centering
\includegraphics[width=0.70\textwidth]{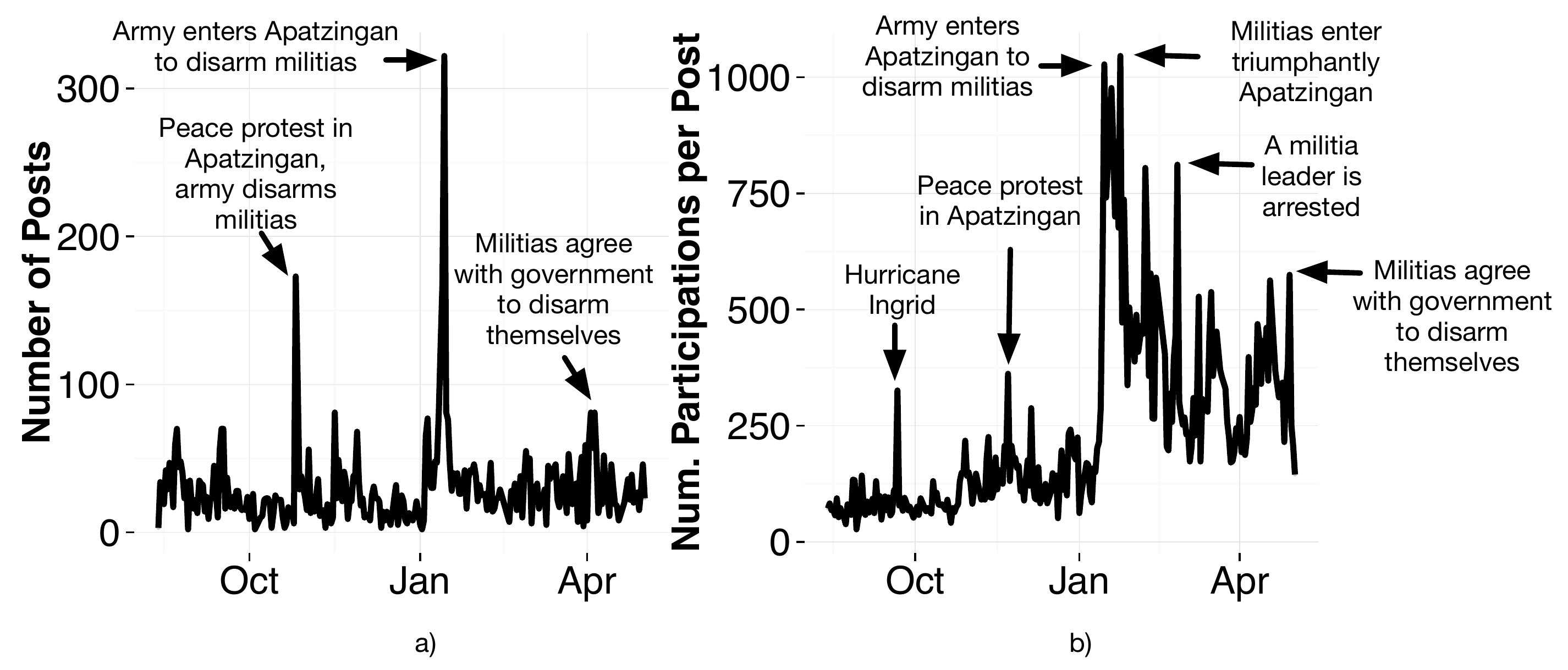}
\caption{
Overview of VXM's posting behavior and audience's activity. In Figure \ref{fig:dailyActivity}a, spikes indicate an increase in the daily number of posts;  in Figure \ref{fig:dailyActivity}b, they show an increase in the daily number of audience likes, comments, and shares per post. Together, VXM and its audience appear to signal when there are major events in Michoac\'{a}n.}
\label{fig:dailyActivity}
\end{figure*}

\subsection{Audience Participation Patterns}

Motivated by Van Dijk's work~\cite{van1997discourse}, we categorize VXM's most active members by looking the messages they post on the page. We focus in particular on the number of messages they post, and their length, and public figures and locations they mention. We used a clustering method using all of these features, except for locations which are used simply to contextualize the results. %We examined the behaviors exhibited by the VXM audience, particularly the commenting behavior of the most active audience members (\emph{the outliers}). We focused on comments because they typically take more effort to produce than a {``like''} and can provide clues to how dedicated a person is in a political/social movement \cite{feezell2009facebook}. 
We identified VXM's \emph{highly active participants} by finding those individuals whose number of comments was higher than three times the standard deviation (normal procedure to find outliers). 

We then profiled the behavior of these active participants by analyzing their total number of comments and the type of public figures and organizations they mentioned. For this purpose, we first grouped all of the public figures and organizations in our list based on their affiliation (i.e., type). For instance, \emph{``Mireles''} and \emph{``Papa Smurf''} were grouped together because both are members of self-defense forces. We obtained the affiliations using data from the \emph{Proceso} and {Wikipedia} articles mentioned earlier. We assumed that a public figure could have only one affiliation. Through this process, we identified four main types of public figures and organizations: government, militia, drug cartel, and journalist (news reporters), and classified each of the public figures in our list into one of these types.

Next, we measured how often each highly active participant mentioned each public figure in their comments. We then used a probabilistic model to implicitly calculate the degree to which they mentioned a certain type of public figure:

\begin{equation}
P(o,T)=\sum_{f \in \mathcal{F}}P(o)P(f|o)P(T|f),
\label{eq:interes}
\end{equation}
\noindent
where $P(o,T)$ is the probability an active participant $o$ mentions a certain type of public figures  $T$ given her comments; $P(o)=n^{-1}$ is the probability of selecting an active participant $o$; $n$ is the total number of participants; 
$P(f|o)=m^{-1}$
is the probability of public figure $f$ appearing in the comments of an active participant  $o$; $m$ is the total number of
words that active participant $o$ has used in her comments; $P(T|f)=1$, when the public figure $f$ is of type $T$  or zero otherwise.

For each highly active participant, we calculated their $P(o,T)$ for all types of public figures and formed a vector representing how often the participant mentioned each type. The size of each vector will correspond to the number of types---in this study, the size is four. %These vectors help profile the behavior of the most active audience members in terms of the number of comments and words they produced, and the type of public figures they mentioned in their comments.

We used the vectors to cluster highly active participants and uncover their behavioral patterns. % of the most active audience members. 
For instance, people who mentioned governmental public figures more than they mentioned militias might be clustered together. We use mean shift algorithm \cite{Cheng:1995:MSM:628321.628711} to group together similar vectors and discover clusters of people. We opted to use mean shift algorithm because it is based on a nonparametric density estimation, and therefore we would not need to know the number of clusters beforehand (unlike K-means). Instead, we let mean shift algorithm discover the clusters from our data. The clusters represent the different behavioral patterns that the most-active audience members exhibited.

We then looked at the characteristics of people in each cluster, particularly (1) the number of comments and words used; (2) the percentage of comments that mention certain types of public figures; and (3) the percentage of comments that reference locations. For the last, we identified comments when each person mentioned a city or town in Michoac\'{a}n. We used a
list\footnote{\url{http://cuentame.inegi.org.mx/monografias/informacion/mich/territorio/div\_municipal.aspx?tema=me}} 
from the National Institute of Statistics and Geography to identify comments that referenced such locations.\\\\

\section{Results}

\subsection{RQ1: The VXM Content and Audience Participation}
Our first research question explores the type of content that militia-related groups share online, and how unaffiliated civilians respond to and participate in, or interact with such content. For this purpose, we examined (1) the daily activity of VXM and its audience; (2) the public figures that VXM covered; and (3) the topics VXM discussed with its audience.

{\bf 1. Daily Activity.}  We plotted the daily number of VXM posts (Figure \ref{fig:dailyActivity}a) and audience participation (likes, comments, and shares) per post (\ref{fig:dailyActivity}b). We used this first analysis as an overview of how active the page and audience are.

Figure \ref{fig:dailyActivity} shows that audience activity per post increased over time, especially in 2014, whereas the the administrators' posting behavior stayed relatively constant. However, both audience and VXM administrators spiked in daily activity. For both, most of the peaks corresponded to times when major offline events occurred in Michoac\'{a}n, especially events concerning the militias. For instance, on January 14, 2014, the day the Army entered the town of Apatzing\'{a}n to disarm the self-defense forces, both the number of posts on the page and the number of times the audience participated per post increased.

What is particularly fascinating is that the audience participation also exhibited spikes not related to the administrator's posting behavior. For instance, on February 8, 2014 when the militia took over the Knights Templar headquarters, the audience had one of its largest activity spikes, whereas the administrators did not increase their number of posts during that time. Hence, the audience might help to provide a more complete picture of when important events are transpiring in Michoac\'{a}n. %An area of future investigation is to examine why VXM only increased its number of posts during some of the militia's major events. Could VXM be posting selectively?

%\clearpage
{\bf 2. Public Figures.} To further inspect the type of content VXM shares and how its audience reacts, we analyzed the top 10\% of names VXM mentioned and studied how the audience responded to these references.

\begin{figure}
\centering
\includegraphics[width=0.45\textwidth]{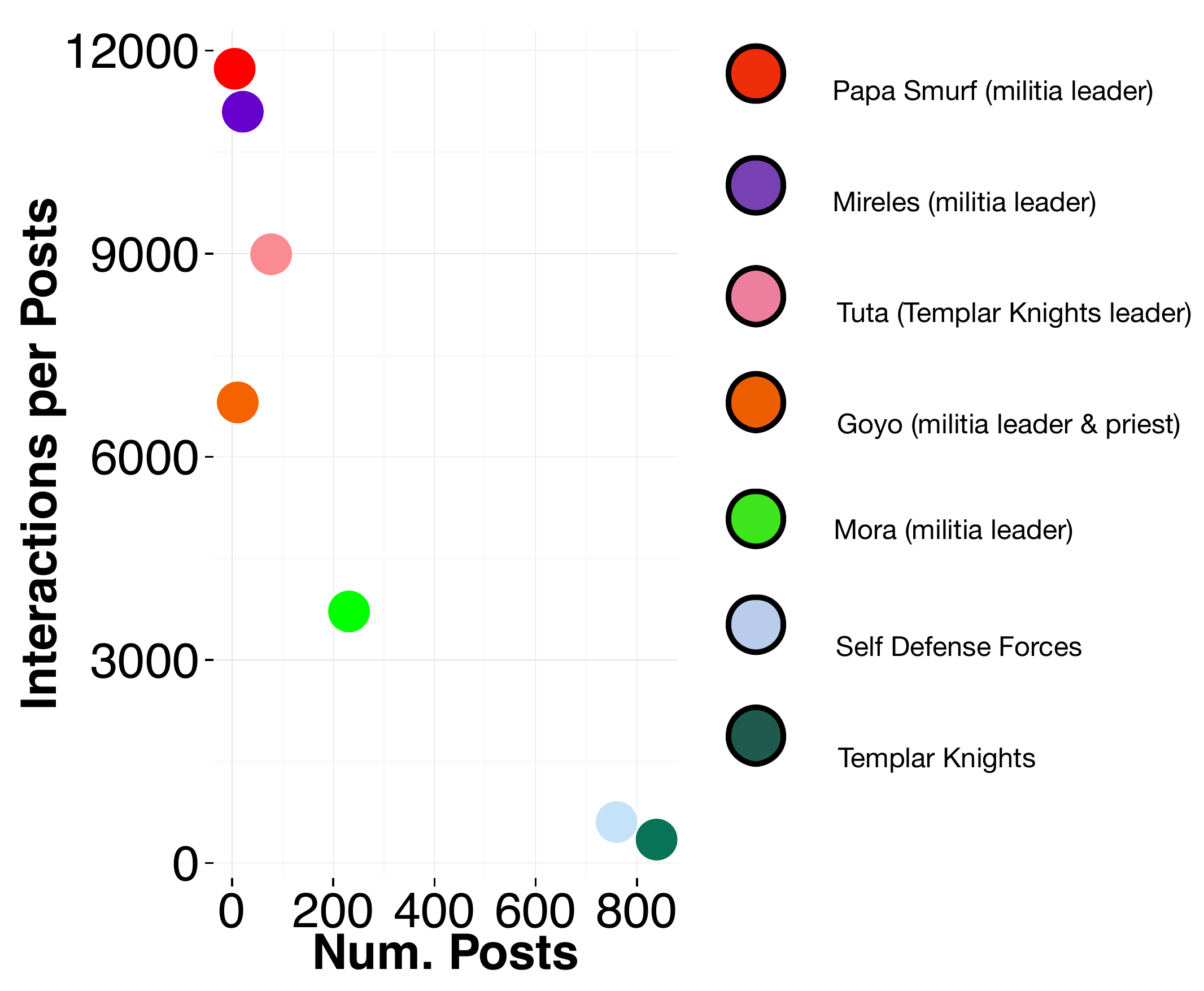}
\caption{Overview of how much VXM mentions certain public figures and the public figures' popularity. The most popular public figures were not necessarily the ones most mentioned by VXM.}

\label{fig:mentionPost}
\end{figure}

Figure \ref{fig:mentionPost} illustrates the total number of posts VXM created about each public figure or organization and compares it with the subject's popularity, measured in terms of the total number of likes, comments, and shares the content received. Each color-coded point represents a public figure or organization. The X axis represents the number of posts about the public figure or organization; and the Y axis, the public figures' popularity. A larger X value represents more mentions by VXM page administrators; a larger Y value represents more popularity with the audience.

Figure \ref{fig:mentionPost}  shows that public figures and organizations that VXM mentioned most often were not actually the most popular with VXM's audience. We also observed that the audience appeared to be more participatory with posts that referenced individuals, rather than organizations. For instance, posts mentioning Knights Templar' leader \emph{``La Tuta'}' and posts mentioning \emph{``Papa Smurf''} were much more popular than posts referencing \emph{``Knights Templar''} or \emph{``Self-Defense Forces.''}

What is surprising about Figure \ref{fig:mentionPost} is that although \emph{Mora} was the militia leader to whom VXM gave the most coverage, \emph{Mora} was not the most popular. The most popular was actually the other militia leader, \emph{Papa Smurf}, whom VXM hardly covered. \emph{Papa Smurf's} popularity might be due to the deal he negotiated with the government in which the self-defense forces officially became a rural police force~\cite{limon}.

It appears again that the audience is helping to provide a more thorough picture of who and what matters in Michoac\'{a}n. Together, VXM administrators and its audience might be framing what important events are taking place in Michoac\'{a}n.

\begin{table*}[htdp]
  \centering \small
  \begin{tabular}{p{9.0cm} p{9.0cm} }
    \hline
   {\bf Topic}&{\bf Sample Post}\\\hline
{\bf Online activity:} information on the nature of the VXM page, pointers to other reliable social media sources, and tips.
& ``If you want to make any complaints about things occurring in your town, try not to publish it on the Facebook wall, send us a private message instead. It is for your own SAFETY. You can also create a fake profile and send to us the complaint, although we will never share your identity.''\\\hline
{\bf Critiques:} criticism of people or organizations, clearly targeted to attack.
&``What do you guys think of this shameless governor, who is a nobody,  who wants to report the brave people that told him his truths? Fausto Vallejo [governor of Michoac\'{a}n] you don't serve as a governor, you are a NOBODY!''\\\hline
{\bf Propaganda:} Patriotic or nationalistic messages used to promote or publicize a political cause or point of view.&No more impunity! we see a Michoac\'{a}n that wants justice that has a hunger for true justice. Only the people can save the people!\\\hline
{\bf News reports:} factual reports, and alerts. Occasionally, the messages include commentary of how the report could impact the reader, or actions the reader could take in relation to the report. 
&``\#SDRLosReyes there is a confrontation, the Knights Templar are attacking in Los Reyes through different entrances. Please alert others who pass those zones. If you have a gun please come help the militias."\\\hline
{\bf Missing people and obituaries:} Reports regarding runaways, possible kidnappings, and deaths.
&``\#MissingPerson \#Morelia, Michoac\'{a}n His name is Jose Manuel Molinero Ruiz."\\\hline
  \end{tabular}
  \caption{Description and sample post of the topics covered by VXM.}
  \label{table:mism}
\end{table*}

\begin{figure}
\centering
\includegraphics[width=0.20\textwidth]{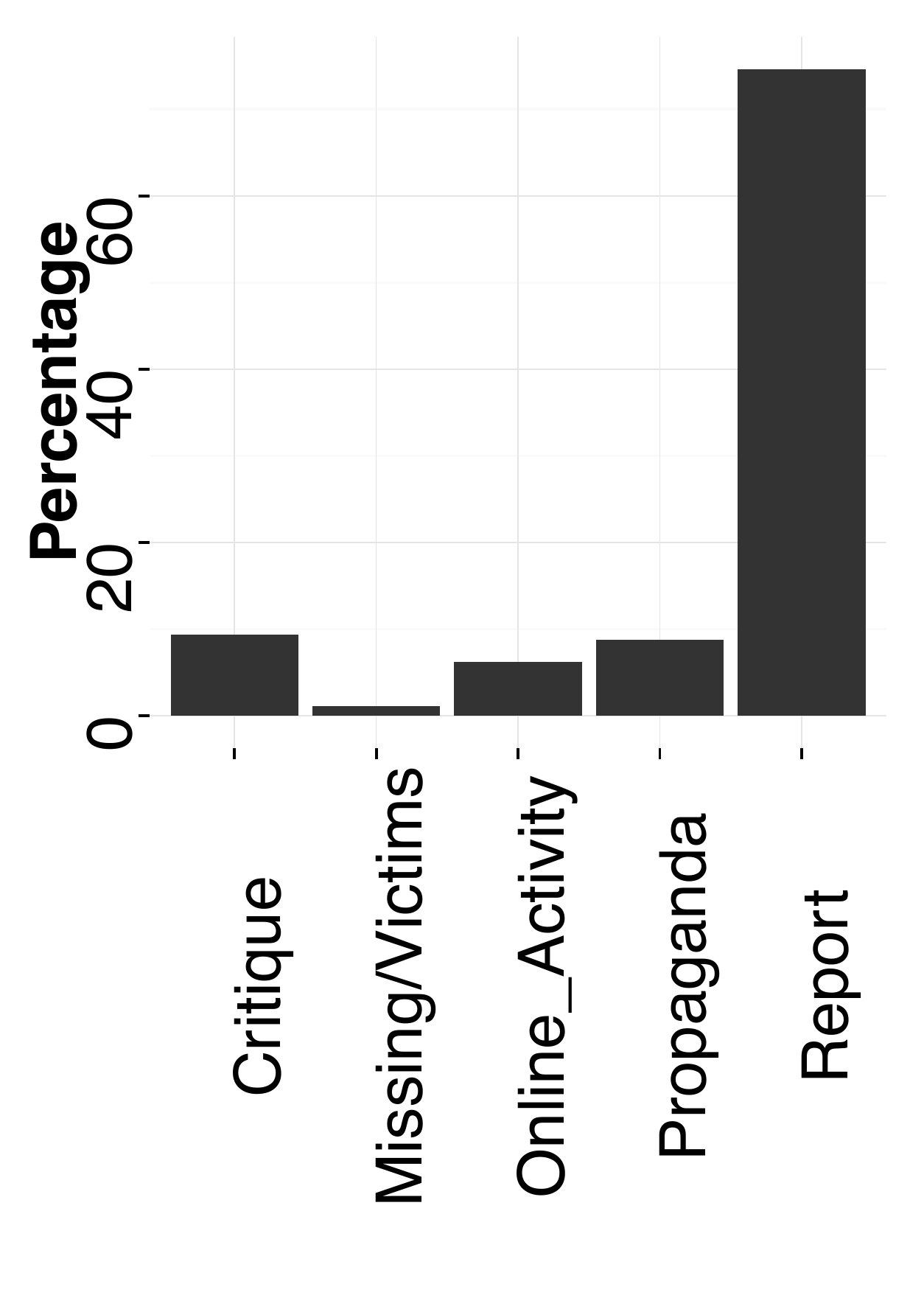}
\caption{Overview of the percentage of posts VXM creates about each topic. VXM primarily provides News Reports. However, occasionally it shares Critiques and Propaganda. It also provides space for its audience by sharing their missing-persons reports, obituaries, and explaining how to keep safe online.}
\label{fig:catp}
\end{figure}

{\bf 3. Topics:} %To further study the type of content that VXM's shares, 
We next examined the type of topics the VXM page presented (Table \ref{table:mism},) and how much coverage the page gave to each (Figure \ref{fig:catp}).

The dominate topic VXM covered was objective news reports. A distinctive trait of these reports was that they usually made requests to the public for some sort of action. For instance, the following VXM post reported a robbery in the town of Los Reyes and asked the public to help identify the criminals: 
\begin{quotation}
\emph{ {``\#LosReyes there was a burglary in the jewelry shop in downtown Los Reyes today at 7 pm. Please check out the YouTube video. If anyone can identify the criminals, please report them via private message.''}}
\end{quotation}

These posts occasionally also included advice for keeping safe: 
\begin{quotation}
\emph{``\#Apatzingan in the dance club Layon the organized crime [Knights Templar] is having an event. WE RECOMMEND NOT ATTENDING. IT IS FOR YOUR OWN SAFETY. The club is full of dangerous people.''}
\end{quotation}

VXM's news reporting appears to be factual and caring at the same time. This type of reporting might help the page have a much deeper connection with its audience and facilitate working together towards a common goal~\cite{jones1964ingratiation}.

Providing news reports aligned with the page's description in its ``About'' section: 
\begin{quotation}
\emph{``Here [at VXM] we provide and follow reports about dangerous situations occurring in Michoac\'{a}n. ...We give coverage to events which take place in the state of Michoac\'{a}n and are of social interest.''} 
\end{quotation}

What is interesting is that VXM also conducted activities not reported in its description, such as sharing propaganda or providing subjective critiques. Of note, VXM incorporated the latter type of content in a subtler way. Only around 10\% of all VXM posts were actually propaganda or critiques.

Communication research has shown that it is hard for a message to reach people who are not already in favor of the views the message presents \cite{cooper1947evasion}. Thus, for speakers who want to plant their own ideas into an audience, the most effective way is to do it subtly, so that the audience is unaware that a report or commentary is actually indoctrination \cite{king1989exposing}. Similarly, if VXM only shared propaganda or critiques, the number of people interested in their content could decrease. Subtly sharing indoctrination types of messages might be an effective technique VXM administrators use to maintain their large audience while still promoting their own ideas or opinions.

Another interesting topic the VXM page covered was online activity. These posts tried to explain the online dynamics of the page, such as what type of online content audiences should expect from the page (e.g., news reports), social media sources the page trusted, and how audiences should conduct themselves on the page. The last item focused primarily on giving tips on how people should present themselves online if they decide to participate in VXM.

This topic resembles a theme seen recently in social media content from terrorist groups \cite{huimas}. Because of the potential danger of being caught, terrorists advise their audiences how to connect to their YouTube videos and present themselves online (e.g., with fake accounts), and which other internet pages are trustworthy allies. Clearly, VXM is not at all related to terrorism. However, there can be value in studying their online activity under the lens of terrorism, especially given the high risks also associated with participating in VXM. The public figures that VXM covered were notoriously dangerous (e.g., drug cartel members). Thus, VXM, just like the terrorists, needed to advise its audience on how to interact in such potentially dangerous online space. Note that the page is an adverse space not necessarily because of its heated discussions, but because drug cartel members could easily eavesdrop on the page's content and violently target audience members whose comments they do not like. It is compelling that VXM makes an effort to explain the page dynamics to its audience and tries to keep them safe.

One topic that was hardly covered by the page (2\% of all posts) but was nonetheless very fascinating was the topic of missing people and obituaries. The page not only provided reports about major events occurring in Michoac\'{a}n, but also provided a space where people could report loved ones that had gone missing, run away, or recently passed away. The VXM page administrators created a space that let people mourn their losses and help each other when they suffered personal family tragedies or emergencies. Thus, VXM appeared socially engaged with its community.\\\\

%What type of militia content attracts the most newcomers,
%i.e., new participants?
\subsection{RQ2: The VXM Newcomers}
Our second research question explored the context under which people started participating in an online space linked to militias. We examined the per-day number of people who participated for the first time with VXM. 

\begin{figure}
\centering
\includegraphics[width=0.40\textwidth]{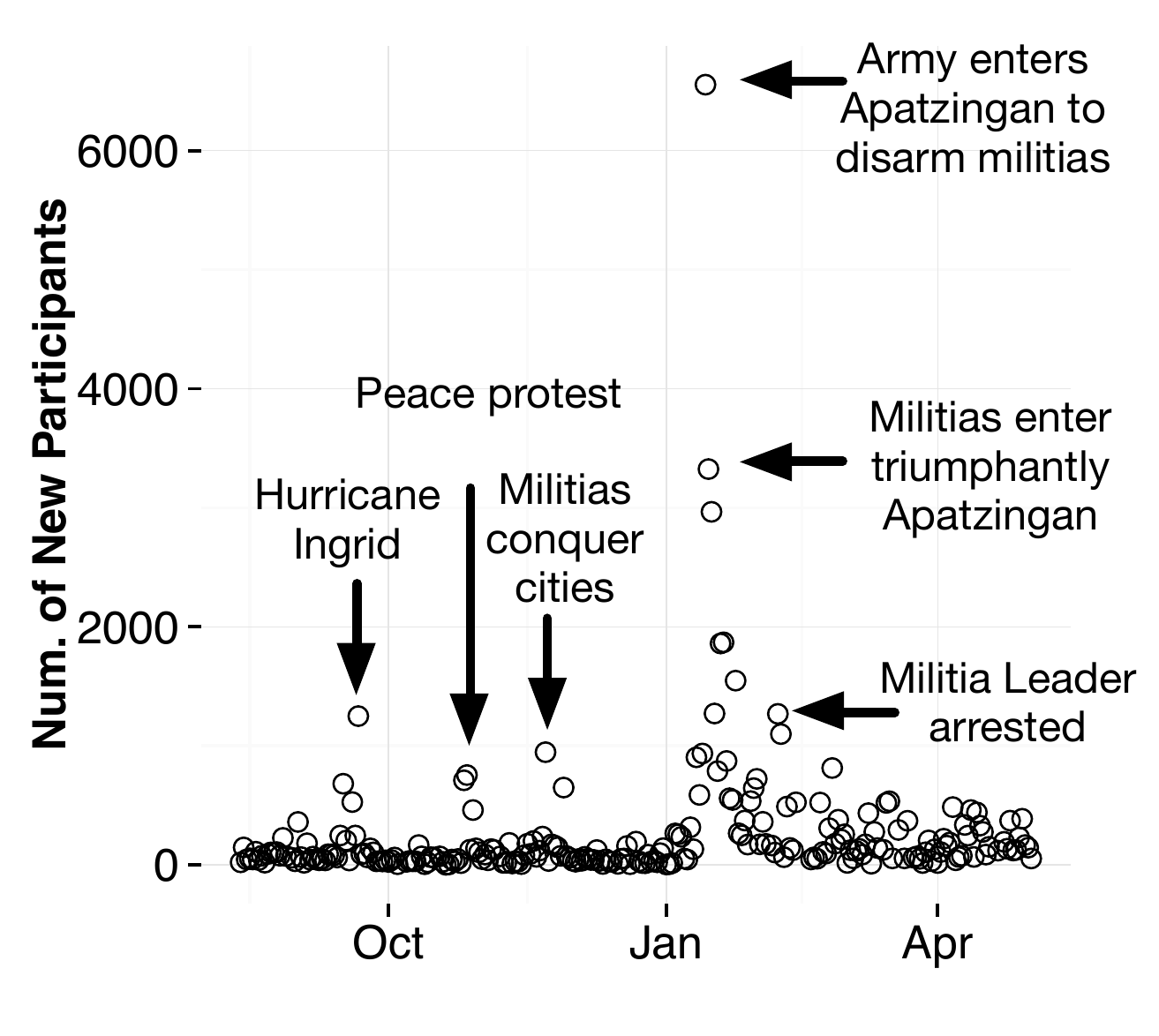}
\caption{Number of daily new participants. In general, the page saw an increased number of new participants whenever there were major offline events.}
\label{fig:firstTime}
\end{figure}

Figure \ref{fig:firstTime} shows number of new participants (newcomers) who either liked or commented on VXM page content. Most of the peaks in the number of daily newcomers correspond to major offline events that occurred in Michoac\'{a}n, especially militia-related events.

For instance, around the time the militias took over the Knights Templar's headquarters on February 8, 2014, the number of newcomers increased. This increase during major offline events might explain the increase in audience participation during the same periods. New, external people suddenly participating in the page could be helping to provide a more complete picture of when there are important offline events in Michoac\'{a}n.

\subsection{RQ3: Behavioral Patterns of the Most Active Audience}
This research question aimed to investigate the behavioral patterns exhibited by VXM's most active audience members. Using a procedure described in ``Audiences Participation Patterns,'' we found that VXM's most active participants (282 individuals) exhibited three distinct behaviors (i.e., we discovered three clusters).

{\bf Cluster A \emph{(Drug Cartel Savvy)}}: 
All individuals in this cluster (33\% of the most active) referenced primarily public figures related to drug cartels. On average, each of these individual's comments referenced drug cartels (67\%), government (12\%), and militias (8\%), with no comments mentioning journalists. This apparent knowledge about drug cartels led us to name people in this cluster the \emph{``Drug Cartel Savvy.''} Drug Cartel Savvy contributed a median of 112 comments; each comment had an average of 36 words (compared with 2 comments, 10 words from average participants, who never mentioned any public figures). Drug Cartel Savvy contributed the largest number and the longest comments of all the behavioral clusters. Sample comments include:
\begin{quotation}
\emph{``Yeah right, as if the Templar [Knights Templar] were so innocent. He [the Knights Templar leader] claims they're not stealing or blackmailing us. I think we need to tell la Tuta [Knights Templar leader] what his boys are up [to]. My mom would say that skunks don't realize how much they stink!''}
\end{quotation}

Forty percent of each of these individual's comments referenced towns in Michoac\'{a}n (average users never mentioned locations). They also referenced the most locations of all the cluster types. A sample comment with locations:
\begin{quotation}
\emph{ ``Right now it is confirmed that Tepalcatepec and Buenavista have those signs [signs that say a town was freed from the Knights Templar]. They look beautiful! I feel pride whenever I see them. God bless the brave people of my neighborhood Loma Blanca! [neighborhood in Tepalcatepec, Michoac\'{a}n].''}
\end{quotation}
\begin{figure}
\centering
\includegraphics[width=0.30\textwidth]{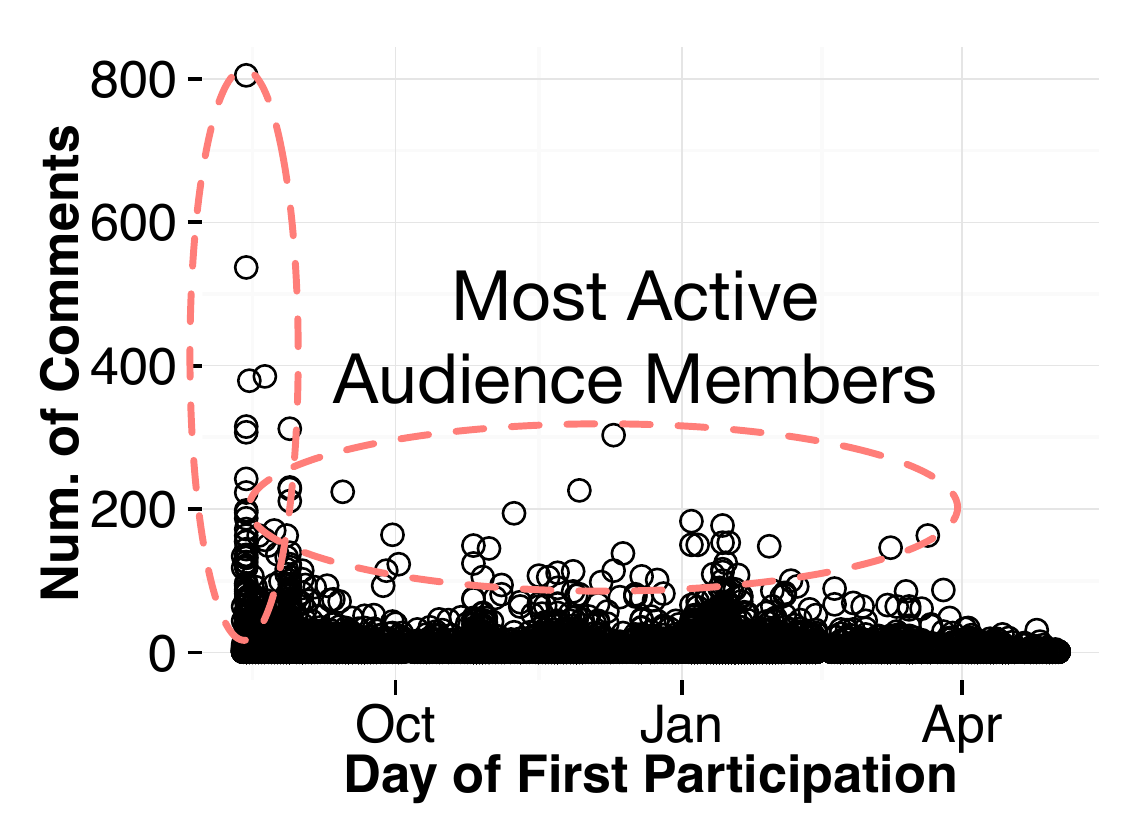}
\caption{Overview of people's start date and the number of comments they generated. The most active audience members started participating either from the start or when the militias had an offline event.}
\label{fig:firstTimePar}
\end{figure}

{\bf Cluster B \emph{(Government Gossipers)}}:
All of these audience members (66\% of the most active) mentioned primarily government figures and related organizations in their comments. On average, each of these individuals' comments referenced the government (28\%), militias (23\%), drug cartels (5\%), journalists (1\%), and towns or cities (4\%). This apparent interest in discussing the government led us to name people in this cluster the \emph{``Government Gossipers.''} Government Gossipers contributed a median of 64 comments; each comment averaged 12 words. Government Gossipers contributed the least content of the three clusters. % in the most-active audience.
However, the majority of the most active audience belonged to this cluster. Sample comments include: 
\begin{quotation}
\emph{``Oh yeah, but fucking Karam [Attorney General] claims he already rescued Michoac\'{a}n. Dirty old liar! He and EPN [President] are a disgrace!''}
\end{quotation}

\begin{quotation}
\emph{``Come on brothers, don't give up! EPN [President] will just have to learn what the people can do together!''}
\end{quotation}

We also found that a few  (1\%) of the Government Gossiper accounts tended to share the exact same comment many times. For instance: 

\begin{quotation}
\emph{``We need to be united as a country, and help the people in Michoac\'{a}n. The government has committed many injustices against the militias, like throwing them in jail. God isn’t with the corrupt. God will not help them.''}
\end{quotation}

\begin{quotation}
\emph{``Even the drug cartels confirm that they are funded by the US. Do you guys think they'll share this on TV?! It is our responsibility to distribute and share this!''}
\end{quotation}

This behavior resembles how people use Twitter in war environments where there is a tendency for redundancy \cite{Mark:2009:RTT:1518701.1518808}.

{\bf Cluster C \emph{(Geographers Michoacanos)}}: 
These audience members (1\% of the most active) did not reference any of the public figures we had on our list. However, they did mention locations. Each person, on average, had some mention or reference to towns and cities in Michoac\'{a}n in 20\% of their comments. We thus decided to name people in this cluster \emph{``Geographers Michoacanos.''}  Although they did not mention the most locations (Drug Cartel Savvy referenced more), they were the only ones who mentioned locations without apparently involving drug cartels, militias, or the government. The Geographers Michoacanos contributed a median of 74 comments; each comment averaged 17 words. A sample comment:\\\\

\begin{quotation}
\emph{``Michoac\'{a}n is home to brave, patriotic, and hard working people. In Nueva Valladolid [former town in Michoac\'{a}n] the first university of all of Latin America was built. The priest Miguel Hidalgo [historical figure] also studied in Nueva Valladolid, today Morelia, [capital of Michoac\'{a}n].''}
\end{quotation}

All of the clusters had audience members with varied start dates. Even the Geographers Michoacanos had accounts that started when the page was founded through the middle of 2014. Figure \ref{fig:firstTimePar} illustrates how much each audience members commented, given their start dates. The X axis shows the date of first comment on VXM; the Y axis shows the number of comments contributed. We highlighted the audience members we considered for this analysis. Note that the majority of the most active participants appear to have started contributing either when the page was first founded or during a major offline event.

\begin{figure}
\centering
\includegraphics[width=0.40\textwidth]{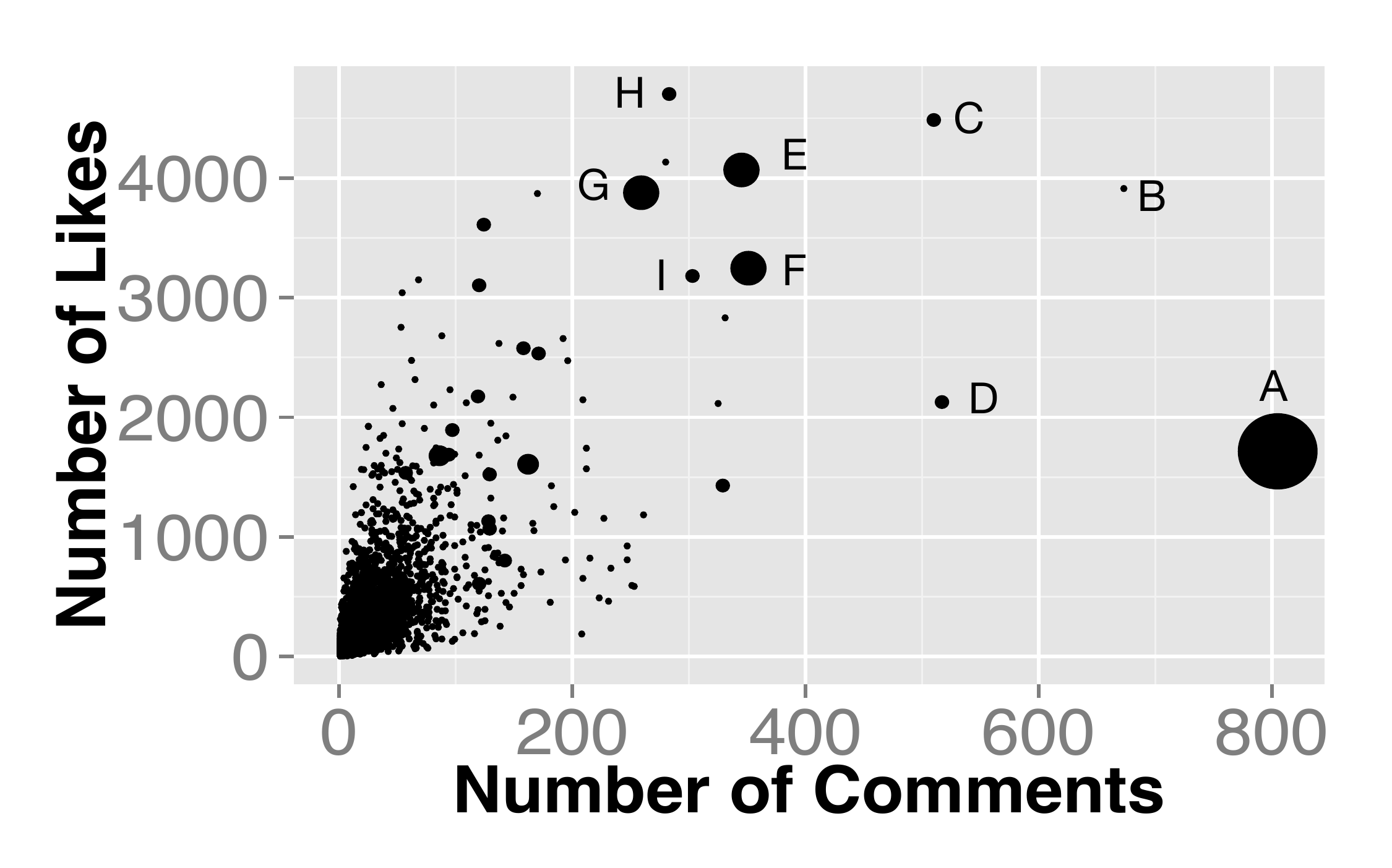}
\caption{Number of likes and comments for all posts, size of a point is directly correlated with the number of shares of a post. The most popular content was published during major offline events of the militias.}
\label{fig:contentPop}
\end{figure}

\begin{table}[htdp]
  \centering
  \small
 \begin{tabular}{ p{.4cm} p{8.0cm} }
    \hline
   {\bf Label}&{\bf  Post Description}\\ \hline
A& { Video of a solider killing an unarmed civilian who supported the militias. } (Jan 15)\\\hline
B& { Post announcing that the militias took over the Knights Templar headquarters.} (Feb 8)\\\hline
C& { Post describing the type of hardships the militias and people in Michoac\'{a}n have endured against the Knights Templar.} (Jan 15)\\\hline
D& { Video of a corrupt police officer.} (Feb 8)\\\hline
E& { Video explaining the militias' motives for fighting.} (Jan 19)\\\hline
F& { Video of  Father Goyo (priest and militia member) justifying the militias' actions. (Jan 14)}\\\hline
G& { Video of Michoac\'{a}n's youth explaining they are tired of living in the violence generated by the Knights Templar.} ( Feb 6)\\\hline
H& { Photo of Father Goyo asking civilians to take action against the drug cartels}. (Feb 8)\\\hline
I& { Photo of Father Goyo crying because the government wants to disarm the militias}. (Jan 18)\\\hline
  \end{tabular}
  \caption{Characteristics of the most popular content. The most popular content was published during major offline events of the militias.}
  \label{table:outliers}
\end{table}

\subsection{RQ4: Content Popularity}
This research question examined some traits of the content that became the most popular in militia-affiliated online spaces; that is, content that received the most number of likes, comments, and shares. Figure \ref{fig:contentPop} illustrates the number of likes, comments, and shares of each VXM post. The X axis shows the number of comments the post received; the Y axis, the number of likes. The size of the circle relates to the number of shares; the larger the circle, the more the post was reshared. We defined VXM's most popular content as the posts whose total number of likes, comments, and shares deviated by three times the standard deviation. We labeled the most popular posts alphabetically and examined a few of their characteristics (Table \ref{table:outliers}).

In general, VXM's most popular content emerged during the militias' most important offline events. For instance, over half of the most popular content was created around January 15, 2014, the day the self-defense militias had their biggest confrontation with the government, who wanted to disarm them. The other half of the most popular content was created around February 8, 2014, the day the militias triumphantly entered Apatzing\'{a}n, a town that had been considered to be the Knights Templar headquarters.

The high number of likes, comments, and shares exhibited during these periods might be due to the increase in newcomers (see Figure  \ref{fig:firstTime}). Content published during major offline events is likely seen by a larger audience and can become popular more easily.

\section{Discussion}

We discuss our results using the previously described themes that tend to appear across all social movements: framing processes, mobilizing structures, and opportunity structures~\cite{kelly2006protest,mcadam1996comparative}.

{\bf Framing Processes.}
Through our case study, we observed that VXM page administrators appeared to: (1) structure and narrate meaning for particular events, public figures, and social life; (2) diagnose what things might need alteration in Michoac\'{a}n; and (3) propose solutions to the diagnosed problems and, in the process, specify what people can do.

The VXM page administrators are likely framing a shared understanding of the situation in Michoac\'{a}n with their audience. Such narrative has also been found in the online activity of US militias \cite{benford2000framing}. However, what is particularly fascinating is that the audience appears to be helping to construct the narrative as well. For instance, %to help narrate the importance of certain towns in Michoac\'{a}n. 
we saw how the audience eventually took hold of the space that VXM had set up and participated copiously whenever there were major offline events. The audience helped to provide a more complete picture of what was happening in Michoac\'{a}n. It was not necessary for VXM to give wide coverage to a certain event or public figure; instead, on its own, the audience adopted the space and participated extensively. The audience appeared to dictate, by itself, what was important.

This behavior has also surfaced in other online spaces. For instance, in ``Hollaback!''~\cite{dimond2013hollaback}, it was the online audience who defined the meaning behind street harassment. In a Facebook group set up to discuss the Coffee Party in the US, certain audience members assumed advocacy roles and extensively supported certain issues \cite{mascaro2011brewing}. Group administrators, in fact, usually provide no input. Traditionally, it had been up to the organization's leaders (e.g., page administrators) or people in power to structure and define what matters. Technology, however, is empowering audiences to establish for themselves what is noteworthy.

{\bf Mobilizing Structures.} We witnessed that VXM was fostering solidarity with its audience by providing safety advice and even assisting them during personal family troubles (e.g., when a loved one passed away). Research on social movements has found that in order to mobilize people and produce action, it is necessary to establish a certain degree of solidarity~\cite{weeber2003militias}. It is likely that the time and space that the VXM administrators are dedicating to their audience is helping them create a highly participatory environment. Note that some of these solidarity conducts resemble online strategies that politicians have adopted to encourage participation from citizens \cite{Robertson:2010:OWP:1858974.1858977}. 

Through our study, we also saw that offline events appeared to be important for mobilizing new participants and popularizing VXM's content. This behavior resembles what has been observed on Twitter  \cite{hughes2009twitter}, where people who joined that service during an emergency event or mass convergence usually became long-term adopters of the technology. Offline events, and the numerous controversies that surround them, might help audiences concentrate on what truly matters to them. This, in turn, is likely helping to increase people's participation in the page. We believe that, in the end, the skillful use of offline events could play a major role in pumping new life into an online group or movement.

{\bf Opportunity Structures.} Our study revealed that some of VXM's most active participants used the page to actively discuss the drug cartels and the government. There are likely several explanations for this behavior. Smelser \cite{smelser1962theory} identified that social strains (i.e., impairments in the relationship between two actors) usually prompt collective action, and the type of strain generally defines the type of collective action that will emerge. In this study, it is likely that these audience members have a strain with the drug cartels or the government  that led them to participate extensively and even exhibit distinct behavioral patterns in VXM. The strains might actually be an opportunity for VXM to recruit highly active participants.

It could also be worth thinking about the role these highly participatory citizens might play in Michoac\'{a}n  or in the country in the long run. Maruyama {et al.} \cite{Maruyama:2014:HMC:2531602.2531719} found that the people who participated on Twitter during a debate were more likely to change their vote in a political election. Their work argued that to take part in an online discussion, people need to process the related information more elaborately; and this could lead to changes in their offline actions. Currently, the offline consequences of citizen participation in militia-affiliated online spaces are unclear. Future work could explore the offline effects that are surfacing from online spaces such as the VXM page.

\section{Design Implications}

Our results raise the question of whether showing the framings  constructed by an audience could benefit social media. Summaries of an audience's constructs could certainly help historians make better sense of a social movement or politically charged event. Audience participation coupled with news reports could provide historians with a more complete picture of when important offline events transpired in a region; and also whether organizations or groups tried to frame reality in a certain way, as their reporting could be compared with the audience's. 

Interfaces highlighting the narrative of an audience could come in different flavors, such as: visualizations of an audience's size and number of participations through time; or visualizations of the audience's interpretations on various matters. Some social media sites, e.g., Facebook or Twitter, already provide tools to visualize the participation of an audience, mainly to improve people's online communications~\cite{bernstein2013quantifying}. Similarly, current research~\cite{savage2014visualizing2,forbes2012visualizing} has explored visualizing the interests and social contexts of audiences to facilitate collaborations. Following these ideas, we believe that interfaces which structure and visualize audiences' framings could facilitate collective action. People could better identify collective thinking about a problem, and could work together towards possible solutions. %Such intrfaces could also certainly help historians make better sense of a social movement. 

Many audience members appeared to have specialized knowledge about organized crime, government, and certain geographical regions.  
There is likely value in designing platforms that can advantageously use audiences' comprehension of the world~\cite{zambranonew}. Perhaps it is about combining citizens' knowledge with the resources from organizations to create smarter  cities~\cite{monroy2013smart}. For example, we could design platforms that integrate the audiences' understanding of the crime in their towns or passion for their neighborhoods~\cite{cranshaw2012livehoods,matias2014newspad,saiphLBS} with government resources to create safer and more lively cities. 

On this last point, note however, that currently individuals and organizations  have to be savvy and learn on their own the best online strategies to spark collaborations or participation from audiences. As designers, it could also be worth thinking about the type of teamwork we want to facilitate. It is unclear whether it is about just building mechanisms through which organizations can donate resources for citizens to use \cite{maruyama2013design}, or if we need to create a more complex platform that intrinsically helps these two players build solidarity with each other. 

Finally, our results found that some of the most active audience members appeared to have social strains against certain organizations. When we design collaborative tools, it could be beneficial to consider these strains (e.g., citizens might not trust the government). We likely need to think about how to create platforms that empower citizens and organizations to collaborate towards improving their cities, while taking into account possible fallouts.

\section{Limitations and Future Work}
The insights this work provides are limited by the methodology and population we studied. For example, the Facebook page we examined focuses on a particular set of militia groups, hence the results might not generalize to all other armed groups. Future work focusing on militia groups in other latitudes could create a body of literature to better understand this phenomenon. 

Also, the methods we used focused on breath rather than depth, as such we do not know much about the identities and motivations of the people behind the Facebook page. Future work could engage in in-depth interviews with the administrators and some of the participants of the page.

Our results also showed that not all major offline events translated into VXM Facebook posts. Previous work has begun an exploration between an individual's online  posts and real-world activities~\cite{kiciman2012omg,mckenzie2013and}. However, the relationship between online posts and real world events in more adverse scenarios has yet to be explored. An understanding of this relationship could impact research areas that range from prediction of a population's activity to crowd-sourced news~\cite{agapieeventful,starCrowd}. 

We found that VXM page administrators conducted different mechanisms to apparently foster solidarity with their audience. Creating solidarity seems to implicitly help VXM to create a highly  participatory online space. Finally, this result suggests that the correlation between the levels of participation of an audience and the solidarity created by page admins warrants deeper investigation.

\section{Acknowledgements}
This work was partially supported by a UC MEXUS-CONACYT fellowship, and by NSF grant IIS-0747520. Special thanks to Tobias Hollerer, Sergio Rodr\'{i}guez, Diego Carmona, Vladimir Beciez and the anonymous CSCW reviewers who helped us improve the paper.

\balance

% If you want to use smaller typesetting for the reference list,
% uncomment the following line:
 \small
\bibliographystyle{acm-sigchi}
\bibliography{VXM}
\end{document}